\documentclass[12pt]{article}

\usepackage{amsmath,amsthm,amsfonts,amssymb,amscd}
\def\qed{{\unskip\nobreak\hfil\penalty50
\hskip2em\hbox{}\nobreak\hfil$\square$
\parfillskip=0pt \finalhyphendemerits=0\par}\medskip}

\def\ind{{\mathrm{Ind}}}

\newtheorem{theorem}{Theorem}[section]
\newtheorem{lemma}[theorem]{Lemma}
\newtheorem{conjecture}[theorem]{Conjecture}
\newtheorem{corollary}[theorem]{Corollary}
\newtheorem{definition}[theorem]{Definition}

\newtheorem{proposition}[theorem]{Proposition}
\newtheorem{remark}[theorem]{Remark}

\def\res{\!\restriction\!}

\def\A{{\cal A}}

\def\B{{\cal B}}
\def\C{{\cal C}}
\def\D{{\cal D}}

\def\PI{{\cal PI}}
\def\PC{{\cal PC}}

\renewcommand{\qed}{\ \hfill $\blacksquare$}

\newcommand{\bdef}{\begin{definition}}
\newcommand{\blem}{\begin{lemma}}
\newcommand{\bprop}{\begin{proposition}}
\newcommand{\bthm}{\begin{theorem}}
\newcommand{\bcor}{\begin{corollary}}
\newcommand{\bconj}{\begin{conjecture}}
\newcommand{\ben}{\begin{equation}}
\newcommand{\een}{\end{equation}}

\newcommand{\ede}{\end{definition}}
\newcommand{\elem}{\end{lemma}}
\newcommand{\eprop}{\end{proposition}}
\newcommand{\ethm}{\end{theorem}}
\newcommand{\ecor}{\end{corollary}}
\newcommand{\econj}{\end{conjecture}}
\newcommand{\brem}{\begin{remark}}
\newcommand{\erem}{\end{remark}}

\newcommand{\ba}{\begin{array}}
\newcommand{\ea}{\end{array}}
\newcommand{\bea}{\begin{eqnarray}}

\title{\huge On Relative Entropy and Global Index}
\author{
{\sc Feng Xu}\footnote{Supported in part by NSF grant DMS-1764157.}\\
Department of Mathematics\\
University of California at Riverside\\
Riverside, CA 92521\\
E-mail: {\tt xufeng@math.ucr.edu}}

\begin{document}
\date{}
\maketitle

\begin{abstract}
Certain duality of relative entropy can fail for chiral conformal
net with nontrivial representations. In this paper we quantify such
statement by defining a quantity which measures the failure of such
duality, and identify this quantity with relative entropy and global
index associated with multi-interval subfactors for a large class of
conformal nets. In particular we show that the duality holds for a
large class of conformal nets if and only if they are holomorphic.
The same argument also applies to  CFT in two dimensions. In
particular we show that the duality holds for a large class of  CFT
in two dimensions if and only if they  are modular invariant. We
also obtain various limiting properties of relative entropies which
naturally follow from our formula.

\end{abstract}

\newpage

\section{Introduction}
In the last few years there has been an enormous amount of work by
physicists concerning entanglement entropies in QFT, motivated by
the connections with condensed matter physics, black holes, etc.;
see the references in \cite{Hollands} for a partial list of
references.    See \cite{Hol}, \cite{Lon}, \cite{L2}, \cite{LXc},
\cite{LXr} , \cite{OT},\cite{Wit} and \cite{X1024}  for a partial
list of recent mathematical work.

This paper is motivated by a very simple fact about von Neumann
entropy. In finite dimensional case the von Neumann entropy of  a
pure state for a matrix algebra $M$ and its commutant $M'$ are
equal, a simple exercise in linear algebra. In the case of conformal
net the algebra $M$ is replaced by the algebra of observables
localized on disjoint union of intervals $I$ denoted by $\A(I).$ The
vacuum state is a pure state.  Hence one may expect that the von
Neumann entropy of vacuum state for  $\A(I)$ and its commutant are
equal. But for type $III$ factors von Neumann entropy is always
infinity so this is not very interesting. By the work of \cite{Cas}
and \cite{LXr} one can define a regularized von Neumann entropy (cf.
Def. \ref{regularized}) for $\A(I)$ , denoted by $G(I)$, which is
finite but not positive, yet verifies equations similar to von
Neumann entropy in the finite dimensional case. When the global
dimension of $\A$ is one, $\A(I)' = \A(I'),$ one can therefore ask
if the regularized von Neumann entropy for $\A(I)$ and  $\A(I)' =
\A(I')$ is the same. This is what we called a duality relation.

It was observed in \S3 of \cite{LXr} that  the regularized  von
Neumann entropy for $\A(I)$ and  $ \A(I')$ are different when the
global dimension of $\A$ is greater than one, and it  is natural to
conjecture that duality relation above   holds if  and only if the
conformal net has global index equal to $1$. The only currently
known example that verify such a relation is the free fermion net
for which we have explicit formulas for mutual information in
general as in \cite{LXr}. One of the goals of this paper is to prove
that this conjecture is true for a large class chiral CFT (Cor.
\ref{cor3}) and also CFT  in two dimensions which are modular
invariant (Cor. \ref{cor22}). For an example, it follows from Cor.
\ref{cor3} that such duality relation is true for  conformal nets
associated with any even positive unimodular lattices. The number of
such lattices grow very fast as their rank increase.

To prove such results we are led to consider a quantity called {\bf
deficit},  which is simply the difference $D_\A(I)=G(I)-G(I')$, and
conjecture (cf. \ref{conj}) that  $D_\A(I)$ is equal to  another
quantity $\hat{D}_\A$ which is defined by using the data associated
with the inclusion $\A(I)\subset \A(I')'$ (cf. \cite{KLM}).  Our key
observation is Th. \ref{main} that   $D_\A(I)- \hat{D}_\A(I)$ remain
the same for a pair of conformal nets $\A\subset \B$ with finite
index. Recall that $D_\A(I)- \hat{D}_\A(I)$ for free fermion nets
can be verified by explicit formulas of \cite{LXr}. It follows that
any conformal net $A$ that is {\bf chain related} to free fermion
net $\A_r$, i.e., there exists a sequence of conformal nets
$\B_1,..., \B_n$ such that $\B_1 =\A, \B_n= \A_r$ and either $\B_i
\subset \B_{i+1}$ or $\B_{i+1} \subset \B_{i}, 1\leq i\leq n-1,$ and
all inclusions are of finite index must verify our conjecture (cf.
Cor. \ref{cor1} and Cor. \ref{cor11}).

To give the reader an idea what kind of equalities  are proved in
this paper let us consider a special case of Cor. \ref{cor1} for a
conformal net $\A$ that is chain related to free fermion net $\A_r.$
Then for $I=I_1\cup I_2, I'=J_1\cup J_2$ we have
$$
S(\omega, \omega_{J_1}\otimes  \omega_{J_2}) -S(\omega,
\omega_{I_1}\otimes  \omega_{I_2}) - \frac{c}{6} \ln \eta =S(\omega,
\omega F_I) - \frac{1}{2} \ln \mu_\A
$$

where $S$ is the relative entropy, $\omega$ is the vacuum state, $c$
is the central charge, $\mu_\A$ is the global index of $\A$, $\eta =
\frac {r_{J_1}r_{J_2}}{r_{I_1}r_{I_2}}$ is a cross ratio, and $F_I:
\A(J_1\cup J_2)'\rightarrow \A(I_1\cup I_2)$ is the conditional
expectation. Previously relations among relative entropies, central
charge and global index are given in asymptotic form in Th. 4.2 of
\cite{LXr}. The above relation is an identity. The duality condition
as described above holds when the righthand side is $0$.

The rest of this paper is as follows: In \S1 after introducing
relative entropy, spatial derivatives,  index for general von
Neumann algebras,  we prove a property of relative entropy
\ref{prop1} which is motivated by our conjecture above. In \S2 we
consider chiral conformal net. We first define a quantity $\D$ which
is  Deficit to measure the failure of duality and we prove our main
theorem  Th. \ref{main}. We deduce Cor. \ref{cor1}, Cor. \ref{cor2}
as consequences of  Th. \ref{main}. In sections \ref{continuous} and
\ref{singular} we apply  Th. \ref{main} to study a number of natural
problems on relative entropy.

In \S3 we consider the two dimensional  CFT cases while essentially
all results of \S2 hold with small modifications.

\section{Preliminaries}\label{prelim}

\subsection{Spatial derivatives, relative entropy and index theory for general subfactors}

Let  $\psi$ be a  normal state on a von Neumann algebra $M$ acting
on a Hilbert space $H$ and $\phi'$ be a  normal faithful state on
the von Neumann algebra $M'$. The Connes spatial derivative, usually
denoted by $\frac{d\psi}{d\phi'}$,  is a positive  operator (cf.
\cite{Con}) . We will use the simplified notation of \cite{OP} and
write $\frac{d\psi}{d\phi'}= \Delta(\frac{ \psi}{\phi'}).$ If $\psi$
is faithful , we have
$$
\Delta(\frac {\psi}{\phi'})^{it} m\Delta(\frac {\psi}{ \phi'})^{-it}
= \sigma^\psi_t(m), \forall m\in M , \Delta(\frac {\psi}{
\phi'})^{it} m\Delta( \frac{\psi}{ \phi'})^{-it} =
\sigma^{\phi'}_{-t}(m), \forall m\in M'
$$
where $\sigma^\psi_t, \sigma^{\phi'}_{-t}$ are modular
automorphisms.

$$
[D\psi_1: \psi_2]_t:=\Delta(\frac {\psi_1}{ \phi'})^{it}
\Delta(\frac {\psi_2}{ \phi'})^{-it}
$$ is independent of the choice of $\phi'$ and is called Connes
cocycle.

Also if $\psi_1\geq \psi_2$ then $$\Delta(\frac {\psi_1}{\phi'})
\geq \Delta(\frac {\psi_2}{ \phi'}).$$ By Page 476 of \cite{Tak}
this is equivalent to $$\frac{1}{1+\Delta(\frac
{\psi_1}{\phi'})}\leq \frac{1}{1+\Delta(\frac {\psi_2}{\phi'})}$$ as
bounded operators.

Suppose $M$ acts on a Hilbert space $H$ and $\omega$ is a vector
state given by $\Omega\in H$. The relative entropy (cf. 5.1 of
\cite{OP}) in this case is $S(\omega, \phi) = -\langle \ln \Delta
(\phi/\omega') \Omega, \Omega\rangle$ where $\omega' $ is the vector
state on $M'$ defined by vector $\Omega$ and $\Delta
(\phi/\omega'):= \frac{d\phi}{d\omega'}$ is Connes spatial
derivative. When $\Omega$ is not in the support of $\phi$ we set
$S(\omega, \phi)= \infty$.


A list of properties of relative entropies that will be used later
can be found in \cite{OP} (cf. Th. 5.3, Th. 5.15 and Cor. 5.12
\cite{OP}):

\bthm\label{515} (1) Let $M$ be a von Neumann algebra and $M_1$ a
von Neumann subalgebra of M. Assume that there exists a faithful
normal conditional expectation $E$ of $M $onto $M_1$. If $\psi$ and
$\omega$  are states of $M_1$ and $M$, respectively, then $S(\omega,
\psi\cdot E) = S(\omega\res M_1, \psi) + S(\omega, \omega\cdot E);$
\par

(2) Let be $M_i$ an increasing net of von Neumann subalgebras of $
M$ with the property $ (\bigcup_i M_i)''=M$. Then $S(\omega_1\res
M_i, \omega_2\res M_i)$ converges to $ S(\omega_1,\omega_2)$ where
$\omega_1, \omega_2$ are two normal states on $M$; \par

(3) Let $\omega$ and  $\omega_1$ be two normal states on a von
Neumann algebra $M$. If $\omega_1\geq \mu\omega,  $ then $S(\omega,
\omega_1) \leq \ln \mu^{-1}$;

(4) Let $\omega$ and  $\phi$ be two normal  states on a von Neumann
algebra $M$, and denote by  $\omega_1$ and  $\phi_1$ the
restrictions of   $\omega$ and  $\phi$ to  a von Neumann subalgebra
$M_1\subset M$ respectively. Then $S(\omega_1, \phi_1)\leq S(\omega,
\phi)$; \par

(5) Let $\phi$ be a normal faithful state on $M_1\otimes M_2.$
Denote by $\phi_i$ the restriction of $\phi$ to $M_i, i=1,2$. Let
$\psi_i$ be  normal faithful states on $M_i, i=1,2$. Then
$$S(\phi, \psi_1\otimes \psi_2)= S(\phi_1,\psi_1) + S(\phi_2,\psi_2)
+ S(\phi,\phi_1\otimes \phi_2)
$$
\ethm
Let $E:M\rightarrow N$ be a normal faithful conditional expectation
onto a subalgebra $N$. $E^{-1}: N'\rightarrow$ is in general an
operator valued weight which verifies the following equation: for
any pair of normal faithful weights $\psi$ on $N$ and $\phi'$ on
$M'$ we have
$$
\Delta(\frac {\psi E}{\phi'}) =\Delta(\frac {\psi }{\phi'E^{-1}})
$$
Kosaki (cf. \cite{Kos}) defined index of $E$, denoted by $\ind E$ to
be  $E^{-1}(1)$.  When $1$ is in the domain of $E^{-1}$, we say that
$E$ has finite index. When both $N, M$ are factors and $E$ has
finite index, we have the (cf. \cite{PP}) Pimsner-Popa inequality
$$E(m)\geq \lambda m, \forall m\in M_+,$$ where
$\lambda= (\ind E)^{-1}.$ The action of the modular group
$\sigma^{\psi E}_t$ on $N'\cap M$ is independent of the choice of
$\psi$. When $E$ is the minimal conditional expectation such action
is trivial on $N'\cap M$. Also the compositions of minimal
conditional expectations are minimal (cf. \cite{LKo}).

\subsection{A result on relative entropy}\label{1.2}

\blem\label{lemma1} Let $A, B$ be positive unbounded operators on a
Hilbert space such that $A\geq B,$ and $\Omega$ is a unit vector
such that $B\Omega = c \Omega$ where $c>0 $ is a constant, $\langle
A\Omega,\Omega\rangle = 1$.  Let $m_A$ be the spectral measure of
$A$ associated with $\Omega$. Then $\int_0^\infty (\ln\lambda)^2 d
m_A(\lambda) <\infty.$\elem

\proof By Page 476 of \cite{Tak} we have that $\frac{1}{1/n + A}\leq
\frac{1}{1/n + B},\forall n>0$ and it follows
$$\int_0^\infty
\frac{1}{1/n+\lambda} d m_A(\lambda) \leq \frac{1}{1/n+c}, \forall
n>0$$

Let $n$ goes to infinity and by Monotone convergence theorem we have
$$
\int_0^\infty \frac{1}{\lambda} d m_A(\lambda) \leq \frac{1}{c},
\forall n>0
$$
We note that $(\ln \lambda)^2 $ is bounded by a constant times
$1/\lambda$ on $(0,1)$, and a constant times $\lambda$ on
$[1,\infty)$. Since by assumption $\int_0^\infty {\lambda} d
m_A(\lambda) = 1,$ we have shown that

$\int_0^1(\ln\lambda)^2 d m_A(\lambda) <\infty,$

$\int_1^\infty (\ln\lambda)^2 d m_A(\lambda) <\infty,$ and the proof
is complete. \qed

\blem\label{lemma2} Let $A$ be a self adjoint operator on a Hilbert
space, and $\Omega $ be a vector in the domain of $A$. Let $f(t)$ be
a strong operator continuous function in a neighborhood of $0$ with
value in the space of bounded operators such that $f(0)$ is
identity. Then
$$
\lim_{t\rightarrow 0} \frac{-i}{t} \langle (e^{itA} -1) f(t)\Omega,
\Omega\rangle = \langle A\Omega, \Omega\rangle
$$\elem
\proof By assumption it is enough to check that
$$
\lim_{t\rightarrow 0} \frac{-i}{t} \langle (e^{itA} -1)
(f(t)-1)\Omega, \Omega\rangle = 0
$$
We note that
$$|| \frac{-i}{t}  (e^{itA} -1)
\Omega||^2 = \int |\frac{1}{t}  (e^{it\lambda} -1)|^2 dm_A(\lambda)
\leq \int |\lambda|^2 dm_A(\lambda) < \infty
$$
$$
||(f(t)-f(0))\omega||
$$ goes to $0$ as $t$ goes to $0$, and the lemma is proved.
\qed

\bprop\label{prop1} Let $M$ be a factor and $\omega$ a normal
faithful state on $M$ acting on the standard representation space
$H$, and $\Omega$ the corresponding vector such that $\langle m
\Omega, \Omega\rangle = \omega (m), \forall m\in M.$ We shall use
the same notation $\omega$ to denote the vector state on $B(H)$ and
its restriction to subalgebras of  $B(H)$.

Let $E_1: M\rightarrow M_1, E_2:M'\rightarrow M_2$ be normal
conditional expectation with finite index, where $M_1, M_2$ are also
factors.  Then
$$
S(\omega, \omega E_1)- S(\omega, \omega E_2)=S(\omega, \omega
E_1E_2^{-1})
$$
and this equation can also be written as
$$
S(\omega, \omega E_1) + S(\omega, \omega E_2^{-1})=S(\omega, \omega
E_1E_2^{-1})
$$

\eprop
 \proof
Ad (1):  By definition we have
$$ S(\omega, \omega E_1)- S(\omega, \omega E_2)=
\lim_{t\rightarrow 0} \frac{-i}{t} \langle (\Delta(\frac{\omega
E_2}{\omega})^{it} -(\Delta(\frac{\omega E_1}{\omega'})^{it})
\Omega, \Omega\rangle
$$

We note that
$$
\Delta(\frac{\omega E_1}{\omega'})^{it} \Omega = \Delta(\frac{\omega
E_1}{\omega'})^{it}  \Delta(\frac{\omega }{\omega'})^{-it}\Omega
=[D\omega E_1: D\omega]_t \Omega
$$

$$
\Delta (\frac{\omega E_1}{\omega E_2})^{it}\Delta (\frac{\omega
E_2}{\omega})^{it}=  \Delta(\frac{\omega E_1}{\omega'})^{it}
\Delta(\frac{\omega }{\omega'})^{-it}
$$

It follows that

$$
S(\omega, \omega E_1)- S(\omega, \omega E_2)= \lim_{t\rightarrow 0}
\frac{-i}{t} \langle (\Delta (\frac{\omega E_1}{\omega
E_2})^{-it}-1)\Delta(\frac{\omega E_1}{\omega'})^{it} \Omega,
\Omega\rangle
$$

Note that $\Delta (\frac{\omega E_1}{\omega E_2})= \Delta
(\frac{\omega E_1 E_2^{-1}}{\omega'}) \geq \mu \Delta (\frac{\omega
}{\omega'})$, for some $\mu >0$. Here the spatial derivative $\Delta
(\frac{\omega }{\omega'})$ is determined by state $\omega$ on $M_2'$
and $M_2$ respectively.

By Lemma \ref{lemma1} and Lemma \ref{lemma2} we have proved the
first equation.  Apply this equation with $E_1$ equal to identity we
get
$$
S(\omega, \omega)- S(\omega,
\omega E_2)=S(\omega, \omega E_2^{-1})
$$
and the second equation follows.

 \qed

It is convenient to formulate the second equation  of the above
Prop. in the following form:

\bcor\label{fun} Let $N_3\subset N_2\subset N_1$ be factors on a
Hilbert space $H$ and $\omega$ is a vector state on $B(H)$ given by
a vector $\Omega\in H$. Let $F_i, N_i\rightarrow N_{i+1}, i=1,2$ be
conditional expectation with finite index. Assume that $\Omega$ is
cyclic and separating for $N_2$. Then
$$
S(\omega, \omega F_2F_1)=  S(\omega, \omega F_2)+ S(\omega, \omega
F_1)
$$
\ecor

\proof This is just a reformulation of the second equation  of Prop.
\ref{prop1} by noting that we can rename  $N_1=M_2', N_2=M, N_3=
M_1, F_1= (\ind E_2)^{-1} E_2^{-1}, F_2= E_1.$ \qed

\begin{remark}\label{additivity}
Under the conditions of the above Cor.  $S(\omega, \omega F)$ is
additive under compositions of conditional expectations, just like
$\ln \ind E$. But of course $S(\omega, \omega F)$ also depends on
the state $\omega.$ This fact plays important role in the proof of
Th. \ref{main} and Th. \ref{equal} in the following.
\end{remark}

\subsection{Chiral CFT case}\label{regularsection}

Let $\A$ be a conformal net (cf. \cite{KLM} and \cite{LXr}) . It is
always split (cf. \cite{Wen} ). Let $\PI$ be the set whose elements
are disjoint union of intervals. If $I$ is an interval on the circle
with two end points $a,b$, $r_I:= |b-a|$ is called the length of
$I$.

For any $I\in \PI$, $\omega_I$ denotes the restriction of $\omega$
to $\A(I)$.  It follows that $\omega_{I_1}\otimes ... \otimes
\omega_{I_n}$ is a normal state on $\A(I)$.

Since we will be concerned with relative entropy of various states,
we introduce some definitions to simplify notations. For $I=I_1\cup
I_2 ...\cup I_n \in \PI$ where $I_i$ are disjoint intervals,
$$\omega^\otimes:= \omega_{I_1}\otimes  \omega_{I_2}\otimes...
\otimes  \omega_{I_n}.$$

A state $\psi$ on $\A(I)$ is said to be {\bf related} to vacuum
state $\omega$ if we can partition $I$ into disjoint union
$I=J_1\cup J_2 ...\cup J_m, J_i\in \PI, 1\leq i\leq m,$ such that
$\psi=\omega_{J_1}\otimes  \omega_{J_2}\otimes... \otimes
\omega_{J_m}.$


We shall consider conformal net whose mutual information for vacuum
state are always finite.

\bdef A conformal net $\A$ is said to have finite mutual information
 if $S(\omega, \omega_I^\otimes) < \infty, \forall I \in \PI$ \ede

Suppose $\A\subset \B$ is an inclusion of conformal nets with finite
index. We shall denote by $E_I: \B(I)\rightarrow \A(I)$ the unique
conditional expectation which preserves the vacuum state when $I$ is
an interval. When $I=I_1\cup I_2 \cup ... \cup I_n$ is a disjoint
union of $n$ intervals, we shall use $E_I$ to denote $E_{I_1}\otimes
... \otimes E_{I_n}$ which is the unique conditional expectation
from $\B(I)$ to  $\A(I)$ which preserves $\omega_{I_1}\otimes ...
\otimes \omega_{I_n}.$

\blem\label{regularlem} (1) If $\A$ has finite mutual information,
then $S(\omega, \psi) < \infty$ for all $\psi$ on $\A(I)$ that is
related to vacuum state $\omega$.

(2) If $\A\subset \B$ and $\B$ has finite mutual information, then
$\A$ also has finite mutual information; \par

(3)   If $\A\subset \B$ has finite index and $\A$ has finite mutual
information, then $\A$ also has finite mutual information.

\elem

\proof

By (5) of Th. \ref{515} we have
$$
S(\omega, \omega_{I\cup J}^\otimes ) = S(\omega, \omega_{I}^\otimes)
+ S(\omega, \omega_{J}^\otimes) +  S(\omega, \omega_{I} \otimes
\omega_{I})
$$
and
$$
S(\omega, \psi_I\otimes \phi_J) = S(\omega, \psi_I) + S(\omega,
\phi_J) +S(\omega, \omega_{I} \otimes \omega_{I})
$$

It follows that any  $S(\omega, \psi)$ can be expressed as linear
combination of  $S(\omega, \omega_{J}^\otimes)$ for suitable
intervals $J\subset I$ and (1) is proved.

(2) follows from definition and monoticity of  relative entropy in
Th. \ref{515}.

By Th. \ref{515} $S_\B(\omega, \omega_{I}^\otimes) - S_\A(\omega,
\omega_{I}^\otimes) = S(\omega, \omega E_I).$ Since $S(\omega,
\omega E_I) \leq \ln (\ind E_I),$ (3) is proved. \qed

It is proved on Page 13 of \cite{X1024} that essentially all known
conformal net (and probably all) has finite mutual information.

A conformal net is called rational if for some $I=I_1\cup I_2,
\bar{I_1}\cap \bar{I_2}=\emptyset $ where the $\A(I)\subset \A(I')'$
has finite index which is called {\it Global index} and is denoted
by $\mu_\A$.

Two conformal nets $A$  and $B$ are said to be  {\bf chain related}
if  there exists a sequence of conformal nets $\B_1,..., \B_n$ such
that $\B_1 =\A, \B_n= \B$ and either $\B_i \subset \B_{i+1}$ or
$\B_{i+1} \subset \B_{i}, 1\leq i\leq n-1,$ and all inclusions are
of finite index. See \S4 of \cite{LXr} for a large class of
conformal nets that are chain related to free fermion nets.

For a conformal net $\A$ with central charge $c$ and finite mutual
information, we define: \bdef\label{regularized} The regularized von
Neumann entropy of vacuum state for $\A(I), I\in \PI$ is defined as
follows: For an interval $I$ we let $G(I):= c/6 \ln r_I,$  $r_I$ is
the length of interval $I$, and
$$G(I_1\cup I_2 \cup ... \cup I_n)= G(I_1)+... +G(I_n)- S(\omega,
\omega_{I_1}\otimes \omega_{I_2}\otimes ... \otimes \omega_{I_n})$$
\ede
 Note that von Neumann entropy for type III factors are always
infinity, and regularized von Neumann entropy as defined are
motivated by the results of \cite{Cas} and \S3 of \cite{LXr}. Note
unlike relative entropy, the regularized von Neumann entropy  is not
always non negative and not invariant under the conformal
transformations on $I$.

When $\mu_\A=1$, $\A(I)= \A(I')' , \forall I\in \PI,$ and the vacuum
state $\omega$ is a pure vector state, we expect that the von
Neumann entropy of $\omega$ for  $\A(I), I\in \PI$ and  $\A(I'),
I\in \PI$ should be the same. Of course both are infinity, but what
is more interesting is to conjecture that $$G(I)= G(I'), \forall
I\in \PI$$ if $\mu_A=1$. In \S3 of \cite{LXr} we have shown that  in
general
$$G(I)\neq G(I')$$ if $\mu_\A >1.$ Hence we expect that
 $$G(I)= G(I'), \forall I\in \PI$$
if and only if  $\mu_\A=1$. At present the only known example which
verifies $\mu_\A=1$ and  $$G(I)= G(I'), \forall I\in \PI$$ is the
free fermion net (cf. \S2 of \cite{LXr}) for which $G(I), \forall
I\in \PI$ is known. To investigate the general cases we define the
following

\bdef\label{deficit} We define the deficit for $\A(I), I\in \PI$ to
be $D_\A(I):= G_\A(I)-G_\A(I').$ \ede

Let $F_I: \A(I')'\rightarrow \A(I)$ be the condition expectation of
index $\mu_\A^{n-1}$ (cf. \cite{KLM}). When there are a pair of nets
involved we shall use the notation $F_{I,\A}$ to avoid confusions.

\bdef\label{deficit2} Let  $I\in \PI$ be  a disjoint union of $n$
intervals, define
$$\hat{D}_\A(I):= S(\omega,\omega F_I) - \frac{n-1}{2}\ln \mu_\A.$$
\ede

The main conjecture of this paper is

\bconj\label{conj} For a rational conformal net
$$D_\A(I)=\hat{D}_\A(I)$$
\econj

Note that when $\mu_\A=1,$ the above conjecture implies that
$$G(I)= G(I'), \forall I\in \PI$$

Suppose $\A\subset \B$ is an inclusion of conformal nets with finite
index. Recall that  $E_I: \B(I)\rightarrow \A(I)$ is the unique
conditional expectation which preserves the vacuum state when $I$ is
an interval. When $I=I_1\cup I_2 \cup ... \cup I_n$ is a disjoint
union of $n$ intervals,  $E_I$  denotes $E_{I_1}\otimes ... \otimes
E_{I_n}$ which is the unique conditional expectation from $\B(I)$ to
$\A(I)$ which preserves $\omega_{I_1}\otimes ... \otimes
\omega_{I_n}.$

We will prove Conj. \ref{conj} for a large class of conformal nets.
The idea is the following : Since we have an important example of
free fermion net $\A_r$ for which we already know
$$D_{\A_r}(I)=\hat{D}_{\A_r}(I)$$, and there are many conformal nets
that are chain related to $\A_r$, if we can show that for a pair of
conformal nets $\A\subset \B$ with finite index that
$$D_\A(I)-\hat{D}_\A(I) = D_\B(I)-\hat{D}_\B(I)$$, then it follows
that Conj. \ref{conj} is true for  conformal nets that are chain
related to $\A_r$.  To state the theorem in more general terms, we
note that assuming that all the quantities involved on the left hand
side are finite, then
$$
D_\A(I)-\hat{D}_\A(I) = D_\B(I)-\hat{D}_\B(I)$$ is equivalent to
$$
S(\omega, \omega E_I) - S(\omega, \omega E_{I'}) =\hat{D}_\A(I)-
\hat{D}_\B(I)
$$

Then the following Th. does exactly this:

\bthm\label{main} (1) Let $\A\subset \B$ be rational conformal nets
with finite index, then
$$
S(\omega, \omega E_I) - S(\omega, \omega E_{I'}) =\hat{D}_\A(I)-
\hat{D}_\B(I)
$$
(2) (1) also holds when $\B$ is free fermion net $\A_r$.

\ethm

\proof Fix $I\in \PI$ which is a disjoint union of $n$ intervals.

Ad (1): Let $E:= (\ind E_{I'} \ind
F_{I',\B})^{-1}E_IF_{I',\B}^{-1}E_{I'}^{-1}$ be the condition
expectation from $\A(I')'\rightarrow \A(I).$ Set
$E_1:=E_IF_{I',\B}^{-1}E_{I'}^{-1}.$

Let us compute $S(\omega, \omega E_I)- S(\omega, \omega F_{I',\B})-
S(\omega, \omega E_{I'})$. Note that $\Omega$ is separating and
cyclic for  $\B(I)'.$ By Prop. \ref{prop1} we have
$$S(\omega, \omega
E_I)- S(\omega, \omega F_{I',\B})- S(\omega, \omega E_{I'}) =
S(\omega, \omega E_1)$$

By \S4 of \cite{Kos} and \cite{LKo} $E$ restricts to trace on
$\A(I)'\cap \A(I')'.$ Let $P_\A$ be the projection in  $\A(I)'\cap
\A(I')'$ which projects onto the closure of ${\A(I)\Omega}$. Then we
have
$$
\Delta (\frac{\omega E}{\omega'})^{it} P_\A \Delta (\frac{\omega
E}{\omega'})^{-it} =P_\A ,\forall t
$$
where $\omega'$ is the state on $\A(I')$ given by $\Omega$. It
follows that $\Delta (\frac{\omega E}{\omega'})$ commutes with
$P_\A$. We note that when restricted to $P_\A \A(I')'P_\A,$ $\omega
E$ is given by $E(P_\A) \omega E_{P_\A}$ where $$ E_{P_\A}: P_\A
\A(I')'P_\A \rightarrow P_\A \A(I)$$ is the unique conditional
expectation and can be identified with $F_{I,\A} : \A(I')'
\rightarrow \A(I)$ where the algebras are on $P_A H_\B = H_\A.$ Note
that $E(P_\A)= [\B:\A]^{-1} = \frac{\mu_\B^{1/2}}{\mu_\A^{1/2}}.$
Hence
$$
\langle \ln \Delta (\frac{\omega E}{\omega'}) \Omega, \Omega\rangle
=  \ln E(P_\A) + \langle \ln \Delta (\frac{\omega
E_{P_\A}}{\omega'}) \Omega, \Omega\rangle =\ln E(P_\A)+ \langle \ln
\Delta (\frac{\omega F_{I,\A} }{\omega'}) \Omega, \Omega\rangle
$$

Note that
$$
\ind E_I = (\frac {\mu_\A}{\mu_\B})^{n/2} , \ind F_{I', \B} =
\mu_\B^{n-1}
$$

Putting the above pieces together we have shown that
$$
S(\omega, \omega E_I)- S(\omega, \omega F_{I',\B})- S(\omega, \omega
E_{I'})=S (\omega, \omega F_{I,\A}) - \frac{n-1}{2} (\ln\mu_\A
+\ln\mu_\B)
$$
Finally by  Prop. \ref{prop1} we have
$$
 - S(\omega, \omega F_{I,\B})= S(\omega, \omega) - S(\omega, \omega F_{I,\B}) = S(\omega,
\omega F_{I,\B}^{-1}) = S(\omega, \omega F_{I',\B}) - (n-1) \ln
\mu_\B
$$
and the proof of the theorem is complete.

Ad (2): Note that in this case $F_{I,\B}$ is identity, so we only
need to evaluate
$$
S(\omega, \omega E_I)- S(\omega, \omega E_{I'})$$

Note that  $ E_{I'}^{-1}: \A(I')'\rightarrow \A_r(I')'=
k\A_r(I)k^{-1}$ where $k$ is the Klein transform.  Let us define
$$\hat{E}_I (kak^{-1}) = E_I(a), \forall a\in \A_r(I)$$

Since $k\Omega= \Omega,$ it follows that
$$
\omega(\hat{E}_I (kak^{-1})) = \omega(E_I(a)), \omega (kak^{-1}) =
\omega(a)
$$ and
$ S(\omega, \omega E_I)= S(\omega, \omega \hat{E}_I).$ Hence by (2)
of Prop. \ref{prop1}

$$
S(\omega, \omega E_I)- S(\omega, \omega E_{I'}) = S(\omega, \omega
\hat{E_I})- S(\omega, \omega E_{I'}) = S(\omega, \omega \hat{E_I}
E_{I'}^{-1})
$$

The rest of the proof is the same as in (1) above.

\qed

By Th. \ref{main} we immediately have

\bcor\label{cor1} If $\A$ is chain related to $\A_r$, then Conj.
\ref{conj} is true for $\A$. \ecor

We also have

\bcor\label{cor2} If $\A$ is chain related to $\A_r$, then $D_\A= 0$
if and only if $\mu_\A=1$. \ecor

\proof If $\mu_\A=1$, then $D_\A= 0$ by Cor. \ref{cor1}. Now suppose
that $D_\A( I_1\cup I_2)= 0.$ By (2) of Th. 4.2 in \cite{LXr}, it
follows that $\mu_\A=1.$\qed

\bcor\label{cor3} Conj. \ref{conj} is true for conformal nets
associated with even positive definite lattices. \ecor

\proof

First we prove this for rank one lattices. Let $\A_{U(1)_a}$ be the
conformal net associated with rank one lattice with $a$ a positive
even integer. Denote by $D_1(a):=
D_{\A_{U(1)_a}}(I)-\hat{D}_{\A_{U(1)_a}}(I).$ We prove by induction
on $k$ that
$$D_1(ka)=D_1(a), \forall k\geq 1$$
When $k=1$ this is trivial. Assume the above equation is true for
$k$. Consider the following finite index inclusions:
$$
U(1)_{(k+1)a} \times U(1)_{(k+1)ka} \subset U(1)_{ka}\times U(1)_{a}
$$
where $U(1)_{(k+1)a}$ is diagonally embedded in  $U(1)_{ka}\times
U(1)_{a}$ and its commutant in  $U(1)_{ka}\times U(1)_{a}$  is
$U(1)_{(k+1)ka}.$

By Th. \ref{main} and induction hypothesis we have
$$
2 D_1((k+1)a)= 2 D_1(a)
$$ and it follows by induction we have proved

$$D_1(ka)=D_1(a), \forall k\geq 1.$$
Now from the inclusion
$$ U(1)_2 \times U(1)_2\subset U(1)_1\times U(1)_1$$

and  Th. \ref{main} we conclude that $D_1(2)=0$. It follows that
$D_1(a)=0$ for all even $a$.

Now assume that the Corollary is proved for all rank $k$ lattices.
If $L$ is an even positive definite lattice, choose a nonzero
element $e\in L$ and consider sublattices $L_1= {\Bbb Z} e$ of $L$
and $L_2$ of $L$ which is orthogonal to $L_1$ with rank equal to
$k$. Apply Th. \ref{main} to the finite index inclusions
$$
\A_{L_1} \otimes \A_{L_2} \subset \A_L$$ and induction hypothesis,
we have proved the Corollary. \qed

\subsection{Some continuous properties}\label{continuous}

Let us first fix a rational conformal net $\A$ with finite mutual
information.

 By (2) of Th. \ref{515} relative entropies are continuous
from ``inside". As an application of Th. \ref{main}, we will prove
that relative entropies in  Th. \ref{main} are also continuous from
``outside". First we  have:

\blem\label{SF1}  If $I\subset J, I, J\in \PI,$ then $F_J$ restrict
to $F_I$ on $\A(I)$ and hence $S(\omega, \omega F_I)$ increase with
$I$; \elem

\proof

This  is proved in \S2 of \cite{KLM} for $n=2$, but the same
argument works for any $n$. \qed

\bcor\label{ouside} Let $\A\subset \B$ be as in Th. \ref{main}. Then
$S(\omega, \omega E_I)$ is continuous from  ``outside", i.e., if
$I_n$ is a decreasing sequence of intervals such that $\cap I_n =I,$
and $E_{I'}$ restrict to $E_{I_n'}$, then
$$\lim_{n\rightarrow \infty} S(\omega, \omega E_{I_n}) = S(\omega, \omega
E_I)$$ \ecor

\proof This follows from Th. \ref{main} and Lemma \ref{SF1}. \qed

\subsection{Singular limits}\label{singular}


It is usually an interesting problem to study the limiting
properties of relative entropies when intervals get close together.
One can find such studies in \S3 and \S4 of \cite{LXr}. In the same
spirit we will consider such singular limits  for related entropy
$S(\omega, \omega F_I)$ for a conformal net $\A$.

The following Theorem is a reformulation of   Proposition 3.25 of
\cite{LXr}:

\bthm\label{325} Assume that $M_n$  is an increasing sequence of
factors act on a fixed Hilbert space, $N_n\subset M_n$ are
subfactors  and $\omega$ is a vector state associated with a vector
$\Omega$. Suppose that $E_n: M_n\rightarrow N_n,  n\geq 1 $ is a
sequence of conditional expectations such that when restricting to
$M_n$, $E_{n+1}=E_n, n\geq 1$, and $\ind E_n = \lambda $ is a
positive real number independent of $n$. If strong operator closure
of $\cup_n N_n$ contains $M_1$, then
$$\lim_{n\rightarrow\infty} S(\omega, \omega E_n)= \ln \lambda
$$\ethm
\proof

Set $\phi_n:= \omega E_n.$

It is sufficient to prove the following as in Proposition 3.25 of
\cite{LXr}: Given any $\epsilon>0$, we need to find $e\in M_n$ for
sufficiently large $n$, such that
$$ |\omega (e)-1|< \epsilon, |\omega (e^*)-1|< \epsilon,\ |\omega
(e^*e)-1|< \epsilon,\ |\phi_n (ee^*)-\lambda|< \epsilon \ .
$$

Let $e_1\in M_1$ be the Jones projection for $E_1:  M_1\rightarrow
N_1$, and $v\in N_1$ be the isometry such that $\lambda v^* e_1 v=
1.$ By assumptions  we can find a sequence of elements $e_n\in N_n,
n\geq 2$
 which converges in strong star topology to
$e_1$.  Now choose $ x_n= \lambda^{-1} v^* e_1 e_n v.$ Then $x_n
\rightarrow 1$ in strong star topology , and so $\omega (x_n),
\omega (x_nx_n^*) $ converges to $1$. On the other hand by
definition
$$
E_n (x_n^* x_n) = v^* e_n^* e_n v
$$ converges to $v^*e_1 v = \lambda^{-1}$ strongly. Hence given  any
$\epsilon>0$, we can choose  $n$ sufficiently large such that if we
set $e=x_n^*$, then  $e\in M_n$, and
$$ |\omega (e)-1|< \epsilon, |\omega (e^*)-1|< \epsilon,\ |\omega
(e^*e)-1|< \epsilon,\ |\phi_n (ee^*)-\lambda|< \epsilon \ .
$$
\qed

Let $I=I_1\cup I_2 \cup ... \cup I_n \in \PI$ and $I'=\hat{I}_1\cup
\hat{I_2} \cup ... \cup \hat{I}_n.$ Let us arrange indices such that
$\hat{I}_i$ share end points with $I_i, I_{i+1}, 1\leq i\leq n-1.$
We are interested in shrinking $I'$. Let us first introduce some
terminology. By {\it a contraction} of $I$ along $\hat{I}_1$ we mean
keep $I_1\cup \overline{\hat{I}_1}\cup I_2:= I_{12}$ fixed and let
the length of $\hat{I}_1$ go to $0$. We will use a  sequence
$I_1(k),\hat{I}_1(k), I_2(k)$ such that $\hat{I}_1(k)$ is decreasing
to describe such a process. Such a sequence is called a {\bf
contraction sequence} along $\hat{I}_1$. Let $C_1(I) = I_{12}\cup
I_3 ... \cup I_n \in \PI.$

\bthm\label{equal} Choosing a contracting $I_1(k),\hat{I}_1(k),
I_2(k)$ sequence  along $\hat{I}_1$. Then
$$
\lim_{k\rightarrow \infty} S(\omega,\omega F_{I}) = S(\omega,\omega
F_{C_1(I)}) + \ln \mu
$$
\ethm \proof Observe that when restricting $ F_{C_1(I)}$ to
$\A(I')'$, we get a conditional expectation simply denoted only in
the proof by $F_k: \A(I')'\rightarrow \A(C_1(I))\cap
\A(\hat{I}_1)'.$ Let $E_{k}: \A(C_1(I))\cap
\A(\hat{I}_1)'\rightarrow \A(I)$ be the conditional expectation such
that $E_{k}$ restricts to identity on $\A(I_3\cup ... \cup I_n)$,
and on $\A(I_{12})\cap \A(\hat{I}_1)'$ is the unique conditional
expectation onto $\A(I_1\cup I_2).$ Note that the index of  $E_{k}$
is $\mu$. Notice that $\Omega$ is cyclic and separating for
$\A(C_1(I))\cap \A(\hat{I}_1)'$. By Cor. \ref{fun} we have

$$
S(\omega,  \omega F_{I}) =S(\omega,  \omega E_kF_k) =S(\omega,
\omega E_k) +S(\omega, \omega F_k)
$$

By (2) of Th. \ref{515} we have $\lim_k  S(\omega, \omega
F_k)=S(\omega,\omega F_{C_1(I)}).$ To finish the proof it is
sufficient to show that
$$
\lim_k S(\omega,  \omega E_k)= \ln\mu.$$

 This follows from Th. \ref{325} since $\cup_k I_1(k)\cup I_2(k)$
is equal to $I_{12}$ minus a point.

\qed

We note that we can apply Th. \ref{equal} a few times to shrink
intervals $\hat{I}_2, ..., \hat{I}_{n-1}$ successively. This way we
see that $$ \lim_k S(\omega,\omega F_{I_k}) = \frac{n-1}{2} \ln
\mu_\A
$$
where one take an increasing of disjoint intervals $I_k$, each one
is a disjoint union $n$ intervals   such that $\cup_k I_k$ is equal
to $S^1$ minus finitely many points. This can of course be proved
directly using Th. \ref{325}.

Now consider the case of $\A\subset \B$ with finite index.

\blem\label{shrink2}
 Choosing a contracting $I_1(k),\hat{I}_1(k),
I_2(k)$ sequence  along $\hat{I}_1$. Then
$$
\lim_{k\rightarrow \infty} S(\omega, \omega E_I) = 1/2(\ln\mu_\A-\ln
\mu_\B) + S(\omega, \omega E_{C_1(I)})
$$
\elem

\proof

For the ease of notations we set $\omega_2:=
\omega_{I_3}\otimes...\otimes \omega_{I_n}.$ By (5) of Th. \ref{515}
$$
S(\omega, \omega_{I_1}\otimes  \omega_{I_2}\otimes \omega_2) =
S(\omega, \omega_{I_1}\otimes  \omega_{I_2})+S(\omega, \omega_2) +
S(\omega, \omega_{I_1\cup I_2}\otimes  \omega_{2})
$$

We note that as  $k$ goes to infinity, $I_1\cup I_2$ increase to
$I_{12}$, hence $$\lim_k S(\omega, \omega_{I_1\cup I_2}\otimes
\omega_{2}) = S(\omega, \omega_{I_{12}}\otimes \omega_{2})$$

Hence
$$
 \lim_k S(\omega, \omega E_I)=  \lim_n S(\omega, \omega E_{I_1\cup I_2}) + S(\omega, \omega E_{C_1(I)} )
$$
The lemma now follows from Th. 4.4 of \cite{LXr}. \qed

\bprop\label{se} Let $\A\subset \B$ be as in Th. \ref{main}.
Choosing a contracting $I_1(k),\hat{I}_1(k), I_2(k)$ sequence  along
$\hat{I}_1$. Then
$$
\lim_{k\rightarrow \infty} S(\omega, \omega E_{I'}) =  S(\omega,
\omega E_{C_1(I')})
$$\eprop

This follows from Th. \ref{main}, Th. \ref{equal} and Lemma
\ref{shrink2}. \qed

The above Cor. can be phrased as follows: Let $I_k=I_{1k}\cup I_2
\cup .. \cup I_n \in \PI$ be such that $I_{1k}$ is a decreasing
sequence such that the length of $I_{1k}$  tends to $0$ as $n$ goes
to infinity. Then
$$
\lim_{k\rightarrow \infty} S(\omega, \omega E_{I_k}) =  S(\omega,
\omega E_{ I_2 \cup .. \cup I_n})
$$
It follows that if either $\A$ or $\B$ has the property that $$
\lim_{k\rightarrow \infty} S(\omega, \omega_{I_k}^\otimes) =
S(\omega, \omega_{ I_2 \cup ... \cup I_n}^\otimes)
$$
then the other net also has this property. In particular all
conformal nets that are chain related to free fermion nets have this
property since  free fermion nets verify such property. It will be
interesting to see if this can be proved under more general
conditions.

\section{ CFT in two dimensions}\label{full}

For a formulation of  CFT in two dimensions we refer to \S2 of
\cite{LK} for more details.

A double cone $C$ is defined to be $I\times J$ where $I, J$ are
intervals on the circle $S^1$ , and we consider $C$ to be a subset
of $S^1 \times S^1.$ Denote by $\PC$ the set which consists of
finite disjoint union of double cones. We shall use $C'$ to denote
the casual complement of $\C$.

We will consider the case $\A\subset \B$ where $A (I\times J) =
\A_L(I)\times \A_R(J),$  both $\A_L$ and $\A_R$ are rational, and
$\A\subset \B$ has finite index. Denote by  $c_L,   c_R$ the central
charges of $\A_L$ and $\A_R$.

\bdef\label{regv2}
 For a double cone $C= I\times J$ we let $G(C):= c_L /6 \ln r_I+  c_R /6  \ln r_J,$
and
$$G(C_1\cup C_2 \cup ... \cup C_n)= G(C_1)+... +G(C_n)- S(\omega,
\omega_{C_1}\otimes \omega_{C_2}\otimes ... \otimes \omega_{C_n})$$
\ede

\bdef\label{deficitb} We define the deficit for $\B(C), C\in \PC$ to
be $D_\B(C):= G_\B(C)-G_\B(C').$ \ede

Note that when the two dimensional  net is tensor product
$\A_L\otimes \A_R,$ and $C= C_1\cup C_2 \cup ... \cup C_n, C_i=
I_i\times J_i, 1\leq i\leq n,$ we have
$$ G_{\A_L\otimes \A_R} (C) =  G_{\A_L} (I_1\cup I_2 \cup ... \cup
I_n) + G_{\A_L} (J_1\cup J_2 \cup ... \cup J_n).$$

Let $F_C: \B(C')'\rightarrow \B(C)$ be the condition expectation of
index $\mu_\B^{n-1}$. \bdef\label{deficit22} When $C$ is a disjoint
union of $n$ double cones,  define
$$\hat{D}_\B(C):= S(\omega,\omega F_C) - \frac{n-1}{2}\ln \mu_\B.$$
\ede

The Conj. \ref{conj} for $\B$  is now

\bconj\label{conj2} For a rational two dimensional   conformal net
$$D_\B(C)=\hat{D}_\B(C)$$
\econj

The proof of Th. \ref{main} applies verbatim to the case of  two
dimensional conformal nets $\A\subset \B$, and we have the following

\bthm\label{main2} (1) Let $\A\subset \B$ be rational two
dimensional conformal nets with finite index, then
$$
S(\omega, \omega E_C) - S(\omega, \omega E_{C'}) =\hat{D}_\A(C)-
\hat{D}_\B(C)
$$
\ethm

\bcor\label{cor11} Suppose $\B$ is chain related to $\A_L\otimes
\A_R$, where both  $\A_L$ and  $\A_R$ are chain related to $\A_r$,
then Conj. \ref{conj2} is true for $\B$. \ecor

We also have

\bcor\label{cor22} (1) Suppose $\B$ is chain related to $\A_L\otimes
\A_R$, where both  $\A_L$ and  $\A_R$ are chain related to $\A_r$
then $D_\B= 0$ if and only if $\mu_\B=1$;

(2) Suppose that $\A_L\otimes \A_R\subset \B,$ and  both  $\A_L$ and
$\A_R$ are chain related to $\A_r$, then  $D_\B= 0$ if and only if
$\B$ is modular invariant.\ecor

\proof

The proof of (1) is the same as the proof of  (1) of Cor.
\ref{cor2}. (2) follows from Th. 4.2 of \cite{Mu}. \qed

A large class of examples with $\mu_\B=1$ can be obtained as
follows: take any conformal net $\A$ which is chain related to free
fermion net and take the Longo-Rehren two dimensional net (which
corresponds to identity modular invariant), it follows by the above
corollary that such net verifies  $D_\B= 0$.

\begin{remark}\label{path} The computation of entropies in physics
literature is usually done (cf. \cite{Casfree}) with replica trick
using path integrals, and when the underlying CFT  can be described
by a Lagrangian  it is usually assumed that the CFT is modular
invariant. In cases where such computations are done, one finds that
the deficit vanishes. Hence (2) of the above Cor. is a rigorous
formulation of such intuitions. \end{remark}

Finally we note that the results of sections \ref{continuous} and
\ref{singular} apply to  two dimensional  conformal nets as well,
with essentially the same proof.


{\footnotesize

\end{document}